# Direct observation of Interatomic Coulombic Decay and subsequent ion-atom scattering in helium nanodroplets

F. Wiegandt[1,2], F. Trinter[1,3,4], K. Henrichs[1], D. Metz[1], M. Pitzer[5], M. Waitz[1], E. Jabbour Al Maalouf[6], C. Janke[1], J. Rist[1], N. Wechselberger[1], T. Miteva[7], S. Kazandjian[7], M. Schöffler[1], N. Sisourat[7], T. Jahnke[1], and R. Dörner[1,†]

[1] *Institut für Kernphysik, Goethe Universität, Max-von-Laue-Strasse 1, D-60438 Frankfurt, Germany*

[2] *Clarendon Laboratory, Department of Physics, University of Oxford, OX1 3PU, United Kingdom*

[3] *Deutsches Elektronen-Synchrotron (DESY), FS-PETRA-S, Notkestrasse 85, D-22607 Hamburg, Germany*

[4] *Fritz-Haber-Institut der Max-Planck-Gesellschaft, Molecular Physics, Faradayweg 4, D-14195 Berlin, Germany*

[5] *Department of Chemical and Biological Physics, Weizmann Institute of Science, P.O. Box 26, 76100 Rehovot, Israel*

[6] *Institut für Ionenphysik und Angewandte Physik, Universität Innsbruck, Technikerstrasse 25, 6020 Innsbruck, Austria*

[7] *Sorbonne Université, CNRS, Laboratoire de Chimie Physique Matière et Rayonnement, UMR 7614, F-75005 Paris, France*

## Abstract

We report on the experimental observation of Interatomic Coulombic Decay (ICD) in pure $^4$He nanoclusters of mean sizes between N~5000-30000 and the subsequent scattering of energetic He$^+$ fragments inside the neutral cluster by using Cold Target Recoil Ion Momentum Spectroscopy (COLTRIMS). ICD is induced in He clusters by using VUV light of hν = 67 eV from the BESSY II synchrotron. The electronic decay creates two neighboring ions in the cluster at a well-defined distance. The measured fragment energies and angular correlations show that a main energy loss mechanism of these ions inside the cluster is a single hard binary collision with one atom of the cluster.

---

[†] Electronic address: doerner@atom.uni-frankfurt.de

**Introduction**

Due to their superfluid and inert characteristics, helium nanoclusters are often used in spectroscopy as a cooling matrix for dopant atoms and molecules in order to study their properties at extremely low temperatures. Being superfluid, these nanodroplets have an almost infinite thermal conductivity. When heated, thermal energy of up to several eV (depending on the cluster size) is dissipated by evaporation of neutral helium atoms from the droplet surface until the equilibrium temperature of 0.38 K is reached again [1]. In addition, movement inside superfluid helium is frictionless below Landau's velocity [2,3]. The measurement of this critical velocity in helium nanodroplets has recently been reported in [4]. Helium droplets are transparent in a broad band reaching from far infrared to vacuum ultraviolet (VUV). The binding energy of each He atom in a droplet is about 0.6 meV [5]. The shape of such droplets is spherical or ellipsoid with a center density of $\rho = 21.8$ nm$^{-3}$ dropping towards zero at the surface within a distance of 6 Å [6-8]. The rotational dynamics of molecules in helium has been extensively studied in the past [1, 9-11] establishing helium droplets as a well-suited environment for studying cold foreign neutral species in liquid helium [12] or improving the spectrometric resolution [13]. Nevertheless, recent X-ray diffraction experiments have indicated the existence of quantum vortices in superfluid droplets [14] implying that deposited rotational energy and angular momentum of up to several thousand ℏ can be absorbed in the clusters by the formation of a large number of quantized vortices [15]. The translational dynamics of neutral species in bulk helium has been studied in the past [16] and has recently been confirmed to proceed in nanodroplets comparable to moving macroscopic objects in bulk superfluid helium [4,17-19]. Photodissociation experiments [20,21] involving $CF_3I$ dissolved in helium nanoclusters revealed a considerable loss of kinetic energy of the fragments inside helium clusters which points to a momentum transfer through binary collisions with cluster atoms. The same collisional model was later successfully applied to explain the velocity distribution of ionic photofragments in helium droplets reported in [22,23].

Here we present a novel approach to address the question of how charged particles move and dissipate energy in helium droplets. We create a singly charged excited $He^{+*}$ ion in the droplet by single photon ionization excitation. We then use the subsequent Interatomic Coulombic Decay (ICD) [24-26] as a very efficient way to deposit a second neighboring charge in undoped helium clusters creating a well-defined initial situation. This allows studying the subsequent interaction of the positively charged ions with the neutral cluster. In this electronic decay process, the ion's excitation energy is transferred through the Coulomb interaction to another cluster atom causing emission of an electron (the "ICD electron") from the second atom. The two neighboring ions repel each other giving rise to a kinetic energy as they fly apart. For isolated helium dimers ($He_2$) in the gas phase ICD is well-known and the kinetic energy distribution of the pair of $He^+$ fragment ions has been examined in great detail [27-32]. In helium droplets, ICD of photoexcited $He^{+*}$ ions has recently been observed [23,33]. The fragmentation dynamics upon ionization of doped helium clusters has first been discussed in [34]. Following the creation of $He^+$ ions in helium droplets, fast migration of the electron hole [35,36] may set in which finally stops resulting in the formation of ionic complexes known as "snowballs" [20] suppressing further charge hopping [37]. Elastic scattering of fast ions with neutral cluster atoms is discussed in the literature [20,22] in which a Monte Carlo simulation based on this model was adapted to fit the experimental data. The same collisional model was successfully applied in [23] to explain the kinetic energy distribution of fast $He^+$ ions in helium clusters. Our results show for the first time a direct, energy- and angle-resolved experimental observation of such scattering events in helium nanoclusters. In contrast, the dynamics of electrons in helium droplets, primarily in cases where the electron kinetic energy is below the threshold for electronic excitation of He, and thus not allowing for inelastic scattering as an energy loss mechanism, is much less understood. Experiments on photoionization of molecules dissolved in helium nanodroplets show indeed very different influence of the helium environment on the photoelectron spectra compared to the spectra of bare molecules [6,33,36,38-40]. In our experiment, we found only negligible energy loss of photo- and ICD electrons (Fig. 1).

**Experimental setup**

We create helium nanoclusters in a supersonic expansion of helium cooled down to a temperature of 12 K through a 5 μm nozzle at a pressure of 25 bar. The neighboring $He^+$ ion pairs are then created inside the helium nanoclusters of mean cluster size between N~5000 and 30000 by using photons (hv = 67 eV) from the BESSY II synchrotron. The photon ionizes and excites a single atom of the cluster into the n=2 excited state of $He^+$. On deexcitation to the ground state, excess energy is either emitted by radiative decay or released through ICD. In the latter process, a second $He^+$ ion is formed inside the cluster and the ion pair dissociates due to the strong Coulomb repulsion (Coulomb explosion). Initially, the fragments are emitted back-to-back and their initial kinetic energy is given by the inverse distance of the atom and the ion at the instant of ICD. Due to energy conservation, this distance is encoded in the IC-decay electron energy, as well. The relative angle between the momentum vectors of the fragment ions leaving the cluster and of the ICD electron as well as their kinetic energy is measured by using the COLTRIMS technique [41-43]. The measurements were carried out at the TGM-7 beamline at Helmholtz-Zentrum Berlin.

**Results**

It is well confirmed by experiments on the helium dimer that the excitation energy of the $He^{+*}$(n=2) state results in an excess energy of 16.22 eV which is shared between the ionic fragments and the ICD electron [24]. This constant sum energy leads to a characteristic diagonal feature when plotting the kinetic energy release (KER) of the ions versus the kinetic energy of the ICD electron (Fig. 1). In turn, the observed diagonal in the energy correlation is a clear proof that ICD does occur in the cluster. It is unlikely, that the observed dissociations occur in the gas phase, as at the conditions to which the nozzle was set the fraction of clusters of N<10 is negligible. To enhance the contrast of this feature, we have selected a subset of the experimental data where the two ion momenta are directed back-to-back with similar magnitude. This filters out events where one or both ions have scattered and lost energy in the cluster.

In case we do not use this back-to-back emission filter on the data, we find that the kinetic energy of most He$^+$ fragments measured in this experiment lies below 3 eV indicating a massive loss of kinetic energy inside the helium cluster. In order to gather more insight into the intracluster kinematics of these ions we restrict our investigations to a different subset of the measured data. By selecting events for which an ICD electron at a kinetic energy of 7.86±0.61 eV was detected in coincidence with two ionic fragments we know the initial energy of each of the two ions to be about 4 eV due to the aforementioned sum energy relation. This suppresses the background consisting of helium ions from direct cluster fragmentation upon single ionization or excitation. Figure 2 depicts the corresponding measured correlation of the final ion kinetic energies after the ions have left the cluster. This energy correlation shows two regions. In region I both ions of the pair carry a low kinetic energy of less than 3 eV per ion, whereas in region II at least one of the ions has retained its initial energy of approximately 4 eV. In the latter case, the other ion shows a broad energy distribution spreading from 0 eV to the maximum energy of 4 eV causing horizontal and vertical structures in Fig. 2.

**Discussion**

Examining these features in more detail allows answering our main question: what is the energy loss mechanism of the ion in the cluster? This mechanism becomes obvious by plotting the angle between the ions versus the ion energy of the slower of the two ions as shown in Fig. 3: the correlation between angle and energy looks completely different for events belonging to regions I and II. For region II a very distinct structure closely following the red line is visible. This line shows the correlation one expects for a single, classical, binary elastic collision of the fast particle with an equal mass particle which is initially at rest. Given the low energy of the cold atoms in the droplet this assumption of being at rest, compared to an ion with a kinetic energy of 4 eV, is well justified. The good agreement between the measured data and the prediction of the binary collision model suggests the following scenario subsequent to ICD: the fast ions (approximately 4 eV) are ions originating from Coulomb explosion which directly leave the cluster without scattering. The ICD-induced second ion initially starts back-to-back to this first ion but on its way through the cluster undergoes one single hard binary collision in which it is deflected and loses part of its initial energy.

In contrast, fragment pairs in region I display a perceptively different energy to angular correlation (Fig. 3, bottom). In this region, both ions are detected with low kinetic energy implying that both ions must have undergone considerable loss of energy. At the same time, their kinetic energy stays comparatively constant over a wide angular range from 180° to approximately 90°. In the context of the elastic scattering model, the very low final kinetic energy of approximately 0.25 eV of fragments in region I implies that both fragments must have separately lost energy in one or multiple scattering events. Such multiple scattering of the ions lead to broad energy or angular distributions when the fragment energies are plotted versus the relative planar angle between their momentum vectors as done in Fig. 3. We find no evidence that the post collisional interaction of the ionic fragments inside the cluster plays a significant role in altering the angular distribution. This can be seen for instance in region II where the energy-to-angle correlation of the scattered fragments closely follows the expected distribution for the case that this interaction is not taken into account. Since ICD electrons with a kinetic energy of 8 eV are detected in coincidence with the fragments in region I, the initial kinetic energy of the fragments is known to be 4 eV per fragment due to the aforementioned constant ICD energy. The energy distribution in region I therefore implies that the fragments have lost approx. 94% of their initial kinetic energy either in few elastic scattering events under large scattering angles or in many elastic scattering events under small scattering angle. It has been suggested in [22] that a sequence of elastic ion-atom scattering events under very small scattering angles would result in a friction-like continuous energy loss of the ions while traveling through the cluster. The model applied therein showed that the average number of collisions in helium clusters of sizes similar to the ones in the present experiment ($N \sim 10^4$) is in the order of 30. However, for the present case a continuous energy loss of 3.75 eV in ~30 scattering events would imply that the average scattering angle is approx. 17°. Since only the planar angle is measured, the angular distribution of the fragments is expected to be isotropic in this case and extend from 0° to 180°. In contrast, energy loss in fewer scattering events in the order of ~2-30 would demand the average scattering angle to be in the order of 17-60°. The large average scattering angle would result in a significantly broadened energy distribution extending from 4 eV down to 0.25 eV and in addition to a broad angular distribution extending from 0° to 180° for low values of the kinetic energy. For both scenarios, we found no evidence that the post collision repulsion of the two charged fragments inside the cluster leads to a significant shift of the angular distribution towards 180° as pointed out above. In fact, the energy distribution and angular distribution in region I do not resemble the distributions expected for any of the scenarios outlined so far. Therefore multiple elastic scattering also in form of a friction like energy loss of the fragments in the cluster does not hold as an explanation for the distribution observed in region I.

A more plausible explanation for the angular distribution observed in region I can be taken from [23]: Upon ICD and Coulomb explosion, fast $He^+$ ions hit neighboring cluster atoms in an elastic collision. Depending on the impact parameter, the fragment ion comes to a complete standstill while transferring most of its kinetic energy to the neutral atom in a head-on binary elastic collision. Another possibility in this context is a fast charge transfer between the neutral atom and the $He^+$ ion which may take place as discussed theoretically for He droplets [35] and observed recently in experiments on small He clusters [46]. In both cases a high energetic neutral atom and a low kinetic energy helium ion are created in the vicinity of the other fragment ion created in the initial ICD event. After this collision, Coulomb explosion sets in again at a much larger internuclear distance resulting in the observed lower KER and angular distribution. However, our data does not allow for a conclusion if one of these two mechanisms is predominant. At the same time, we can rule out a larger initial internuclear distance, i.e. ICD between non-nearest neighbors (second-shell ICD) [47], as an explanation for the low fragment energies observed in region I as such IC-decays would result in ICD electrons with much higher kinetic energy than detected in coincidence with the $He^+$ fragments in region I and II.

The present experimental data indicate a considerable interaction of charged atoms with the helium cluster. The data show clear signatures for elastic scattering (region II). These findings can be reconciled considering the droplet size in our experiment. At this size, the surface layer amounts up to about 30% of the volume. We have performed further measurements on the relative contributions in region I as compared to region II (not shown) and found that the relative yield in region I increases with increasing cluster size. Compiling all results, we suggest the following overall scenario: if the ion pair is created inside the droplet [Fig. 2(b)], in the majority of cases both ions are slowed down leading mostly to events in region I. If, however, ICD occurs at the surface, two options are possible: firstly, if the IC-decay occurs between two partners which are located both on the surface, they both can escape without interaction with neutral partners leading to two ions of equal kinetic energy of about 4 eV emitted under an angle of 180° [Fig. 2(c)]. Secondly, if the pair participating in ICD is oriented perpendicularly to the surface [Fig. 2(a)], one ion escapes without energy loss and without scattering. The second ion is shot into the bulk of the droplet and loses its energy mainly in a binary collision. This scenario can further be supported by examining the angular distribution of the ICD electron with respect to the direction of the fast ion [Fig. 4(b)] in comparison to the ICD electron angular distribution occurring for ICD in the dimer [Fig. 4(a)]. For the cluster case, we find a slight suppression for an ICD electron emission in direction of the slower ions in both regions I and II. This is in line with the surface scenario outlined above. The ICD electrons emitted away from the supposed cluster surface (i.e. in the direction of the fast ion) reach the detector, without perturbation. The ICD electrons which initially are emitted into the cluster are, however, partly slowed down and/or even absorbed in the cluster and hence do not fall into the energy region of ICD electrons (7.86±2.00 eV) selected in Fig. 4(b). The cluster thus shadows the electron emission leading to the slight asymmetry observed in Fig. 4(b).

**Conclusion**

In conclusion, Interatomic Coulombic Decay of neighboring helium atoms in superfluid helium nanodroplets of different sizes was experimentally observed and supports the results reported in [23]. We found that the ionic fragments emerging from the dissociation of neighboring cluster atoms strongly interact with the helium environment through elastic scattering. The experimental results validate the applicability of the collisional model for fast moving ions in helium droplets. We observed that approximately 27% of the fragments (Fig. 2, events in region II) are elastically scattered from neutral cluster atoms resulting in a momentum transfer closely correlated to the scattering angle. 73% of the fragments, however, lose almost their entire initial kinetic energy through a friction-like interaction which significantly shifts the fragment energies to lower values but still leads to a strongly non-isotropic angular distribution (Fig. 2, events in region I). Second-shell ICD as a cause for the extremely low fragment energies can be excluded but might be addressed as a subject to future experiments in the field of helium nanodroplets.

**Acknowledgement**

We thank Helmholtz-Zentrum Berlin for the allocation of synchrotron radiation beamtime. The authors' special appreciation is extended in particular to Christian Pettenkofer for the allocation of the TGM-7 beamline at the BESSY II synchrotron. We thank Stephan Denifl for support of this project and advice on the interpretation of the data. The present work is supported by the DFG Research Unit FOR1789. N. S. thanks financial support from Agence Nationale de la Recherche through the program ANR-16-CE29-0016-01.

**References:**

[1] M. Hartmann, R. E. Miller, J. P. Toennies, and A. Vilesov, *Phys. Rev. Lett. 75, 1566 (1995).*

[2] L. Landau, *Journal Of Physics-USSR 5, 71 (1941).*

[3] M. Schlesinger, M. Mudrich, F. Stienkemeier, and W. T. Strunz, Chem. *Phys. Lett. 490, 245 (2010).*

[4] N. B. Brauer, S. Smolarek, E. Loginov, D. Mateo, A. Hernando, M. Pi, M. Barranco, W. J. Buma, and M. Drabbels, *Phys. Rev. Lett. 111, 153002 (2013).*

[5] J. H. Kim, D. C. Peterka, C. C. Wang, and D. M. Neumark, *J. Chem. Phys. 124, 214301 (2006).*

[6] M. Rosenblit and J. Jortner, *J. Chem. Phys. 124, 194505 (2006).*

[7] K. vonHaeften, T. Laarmann, H. Wabnitz, and T. Möller, *Phys. Rev. Lett. 87, 153403 (2001).*

[8] F. Stienkemeier and K. K. Lehmann, *J. Phys. B. 39, R127 (2006).*

[9] S. Grebenev, J. P. Toennies, and A. F. Vilesov, *Science 279, 2083 (1998).*

[10] C. Callegari, A. Conjusteau, I. Reinhard, K. K. Lehmann, and G. Scoles, *J. Chem. Phys. 113, 10535 (2000).*

[11] D. Pentlehner, J. H. Nielsen, A. Slenczka, K. Mølmer, and H. Stapelfeldt, *Phys. Rev. Lett. 110, 093002 (2013).*

[12] K. Nauta and R. E. Miller, *Science 283, 1895 (1999).*

[13] K. Nauta and R. E. Miller, *Science 287, 293 (2000).*

[14] L. Gomez et al., *Science 345, 906 (2014).*

[15] K. K. Lehmann and A. M. Dokter, *Phys. Rev. Lett. 92, 173401 (2004).*

[16] H. Günther, M. Foerste, M. Kunze, G. z. Putlitz, and U. v. Stein, *Z. Phys. B. 101, 613 (1996).*

[17] S. Smolarek, N. B. Brauer, W. J. Buma, and M. Drabbels, *J. Am Chem. Soc. 132, 14086 (2010).*

[18] L. Chen, J. Zhang, W. M. Freund, and W. Kong, *J. Chem. Phys. 143, 044310 (2015).*

[19] F. Filsinger, D.-S. Ahn, G. Meijer, and G. v. Helden, *Phys. Chem. Chem. Phys. 14, 13370*

*(2012).*

[20] A. Braun and M. Drabbels, *Phys. Rev. Lett. 93, 253401 (2004).*

[21] A. Braun and M. Drabbels, *J. Chem. Phys. 127, 114303 (2007).*

[22] D. S. Peterka, J. H. Kim, C. C. Wang, and D. M. Neumark, *J. Phys. Chem. B 110, 19945 (2006).*

[23] M. Shcherbinin, A. C. LaForge, V. Sharma, M. Devetta, R. Richter, R. Moshammer, T. Pfeifer, and M. Mudrich, *Phys. Rev. A 96, 013407 (2017).*

[24] L. S. Cederbaum, J. Zobeley, and F. Tarantelli, *Phys. Rev. Lett. 79, 4778 (1997).*

[25] U. Hergenhahn, *J. Electron Spectrosc. Relat. Phenom. 184, 78 (2011).*

[26] T. Jahnke, *J. Phys. B: At. Mol. Opt. Phys. 48, 082001 (2015).*

[27] T. Havermeier, T. Jahnke, K. Kreidi, R. Wallauer, S. Voss, M. Schöffler, S. Schössler, L. Foucar, N. Neumann, J. Titze, H. Sann, M. Kühnel, J. Voigtsberger, J. H. Morilla, W. Schöllkopf, H. Schmidt-Böcking, R. E. Grisenti, and R. Dörner, *Phys. Rev. Lett. 104, 133401 (2010).*

[28] N. Sisourat, N. V. Kryzhevoi, P. Kolorenč, S. Scheit, T. Jahnke, and L. S. Cederbaum, *Nat. Phys. 6, 508 (2010).*

[29] F. Trinter, J. B. Williams, M. Weller, M. Waitz, M. Pitzer, J. Voigtsberger, C. Schober, G. Kastirke, C. Müller, C. Goihl, P. Burzynski, F. Wiegandt, T. Bauer, R. Wallauer, H. Sann, A. Kalinin, L. Ph. H. Schmidt, M. Schöffler, N. Sisourat, and T. Jahnke, *Phys. Rev. Lett. 111, 093401 (2013).*

[30] J. Titze, M. S. Schöffler, H.-K. Kim, F. Trinter, M. Waitz, J. Voigtsberger, N. Neumann, B. Ulrich, K. Kreidi, R. Wallauer, M. Odenweller, T. Havermeier, S. Schössler, M. Meckel, L. Foucar, T. Jahnke, A. Czasch, L. Ph. H. Schmidt, O. Jagutzki, R. E. Grisenti, H. Schmidt-Böcking, H. J. Lüdde, and R. Dörner, *Phys. Rev. Lett. 106, 033201 (2011).*

[31] H.-K. Kim, H. Gassert, J. N. Titze, M. Waitz, J. Voigtsberger, F. Trinter, J. Becht, A. Kalinin, N. Neumann, C. Zhou, L. Ph. H. Schmidt, O. Jagutzki, A. Czasch, M. Schöffler, H. Merabet, H. Schmidt-Böcking, T. Jahnke, H. J. Lüdde, A. Cassimi, and R. Dörner, *Phys. Rev. A 89, 022704 (2014).*

[32] P. Burzynski, F. Trinter, J. B. Williams, M. Weller, M. Waitz, M. Pitzer, J. Voigtsberger, C. Schober, G. Kastirke, C. Müller, C. Goihl, F. Wiegandt, R. Wallauer, A. Kalinin, L. Ph. H. Schmidt, M. Schöffler, G. Schiwietz, N. Sisourat, T. Jahnke, and R. Dörner, *Phys. Rev. A 90,*

*022515 (2014).*

[33]     Y. Ovcharenko, V. Lyamayev, R. Katzy, M. Devetta, A. LaForge, P. O'Keeffe, O. Plekan, P. Finetti, M. Di Fraia, M. Mudrich, M. Krikunova, P. Piseri, M. Coreno, N. B. Brauer, T. Mazza, S. Stranges, C. Grazioli, R. Richter, K. C. Prince, M. Drabbels, C. Callegari, F. Stienkemeier, and T. Möller, *Phys. Rev. Lett. 112, 073401 (2014).*

[34]     M. Lewerenz, B. Schilling, and J. Toennies, *J. Chem. Phys. 102, 8191 (1995).*

[35]     N. Halberstadt and K. C. Janda, *Chem. Phys. Lett. 282, 409 (1998).*

[36]     D. Buchta, S. R. Krishnan, N. B. Brauer, M. Drabbels, P. O'Keeffe, M. Devetta, M. Di Fraia, C. Callegari, R. Richter, M. Coreno, K. C. Prince, F. Stienkemeier, J. Ullrich, R. Moshammer, and M. Mudrich, *J. Chem. Phys. 139, 084301 (2013).*

[37]     M. Mudrich and F. Stienkemeier, *In. Rev. Phys. Chem. 33, 301 (2014).*

[38]     D. S. Peterka, A. Lindinger, L. Poisson, M. Ahmed, and D. M. Neumark, *Phys. Rev. Lett. 91, 043401 (2003).*

[39]     E. Loginov, D. Rossi, and M. Drabbels, *Phys. Rev. Lett. 95, 163401 (2005).*

[40]     M. Rosenblit and J. Jortner, *Phys. Rev. Lett. 75, 4079 (1995).*

[41]     R. Dörner, V. Mergel, O. Jagutzki, L. Spielberger, J. Ullrich, R. Moshammer, and H. Schmidt-Böcking, *Phys. Rep. 330, 95 (2000).*

[42]     J. Ullrich, R. Moshammer, A. Dorn, R. Dörner, L. Ph. H. Schmidt, and H. Schmidt-Böcking, *Rep. Prog. Phys. 66, 1463 (2003).*

[43]     T. Jahnke, Th. Weber, T. Osipov, A. L. Landers, O. Jagutzki, L. Ph. H. Schmidt, C. L. Cocke, M. H. Prior, H. Schmidt-Böcking, and R. Dörner, *J. Electron Spectroc. Relat. Phenom. 141, 229 (2004).*

[44]     P. Radcliffe, A. Przystawik, T. Diederich, T. Döppner, J. Tiggesbäumker, and K.-H. Meiwes-Broer, *Phys. Rev. Lett. 92, 173403 (2004).*

[45]     A. Conjusteau, Dissertation: Spectroscopy in Helium Nanodroplets: Studying Relaxation Mechanisms in Nature's Most Fascinating Solvent, Princeton University (2002).

[46]     S. Kazandjian, J. Rist, M. Weller, F. Wiegandt, D. Aslitürk, S. Grundmann, M. Kircher, G. Nalin, D. Pitters, I. Vela Pérez, M. Waitz, G. Schiwietz, B. Griffin, J. B. Williams, R. Dörner, M. Schöffler, T. Miteva, F. Trinter, T. Jahnke, and N. Sisourat, *Phys. Rev. A 98, 050701(R) (2018).*

[47] E. Fasshauer, *New J. Phys. 18, 043028 (2016).*

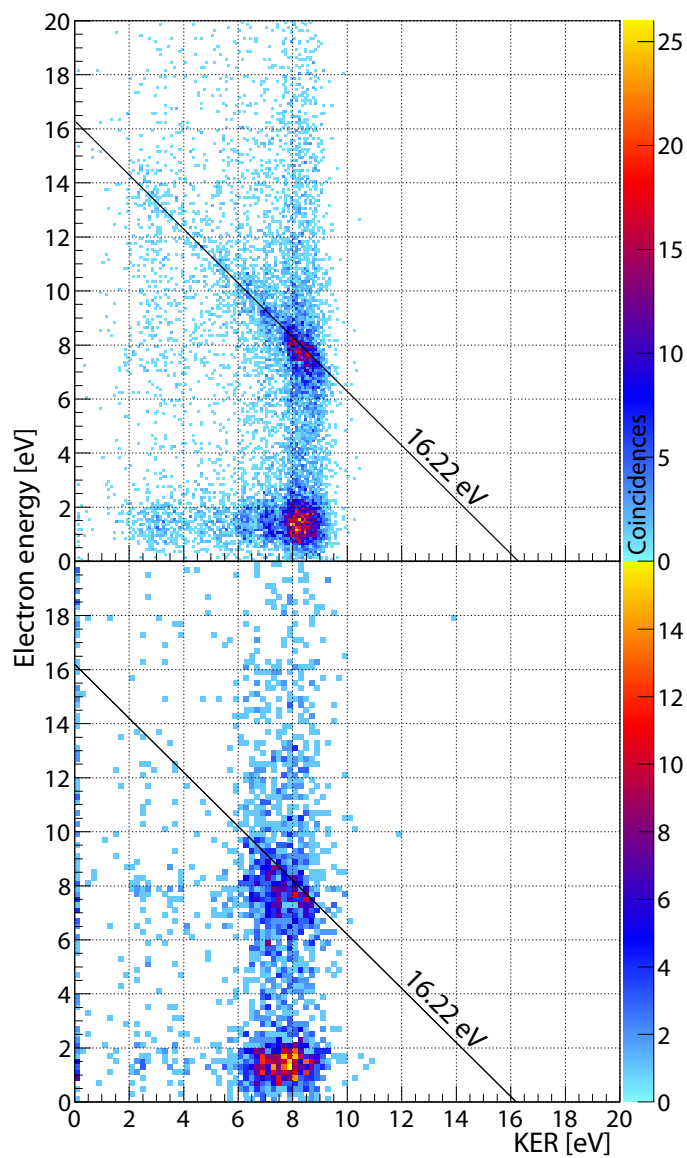

**Figure 1** Joint energy spectrum showing ICD in helium dimers (top) and in N~5000 clusters (bottom). The diagonal features show that the decay energy is shared between the ICD electron and the two He$^+$ ionic fragments.

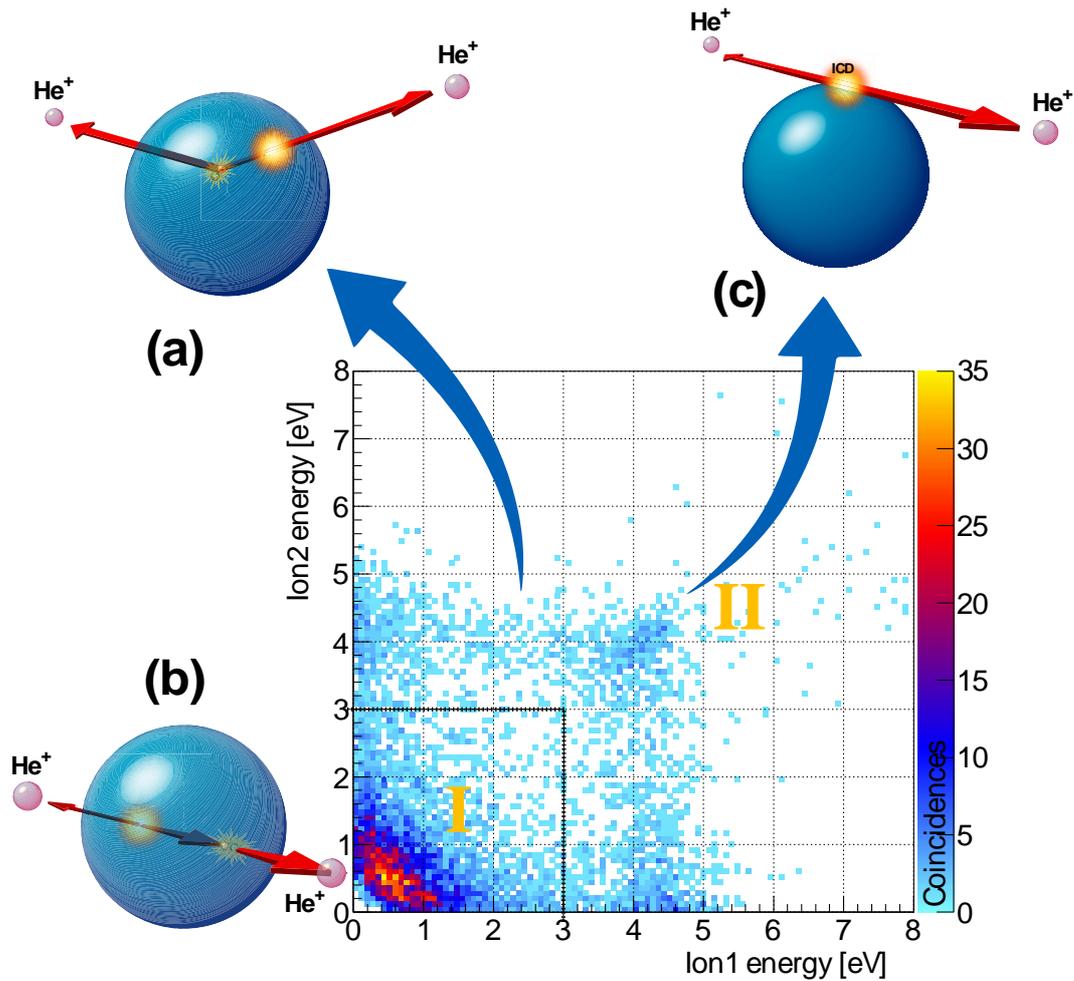

**Figure 2** Joint energy spectrum of both He$^+$ fragments after ICD showing mostly events with fragment energies below 3 eV (region I) and few events with higher particle energies (region II). (a) The ion pair is created near the cluster surface. One ion is shot into the bulk of the droplet and is scattered from a neutral cluster atom. (b) The ion pair is created inside the droplet and is slowed down due to interaction with neutral cluster atoms. (c) Both ions escape tangentially to the cluster surface without energy loss.

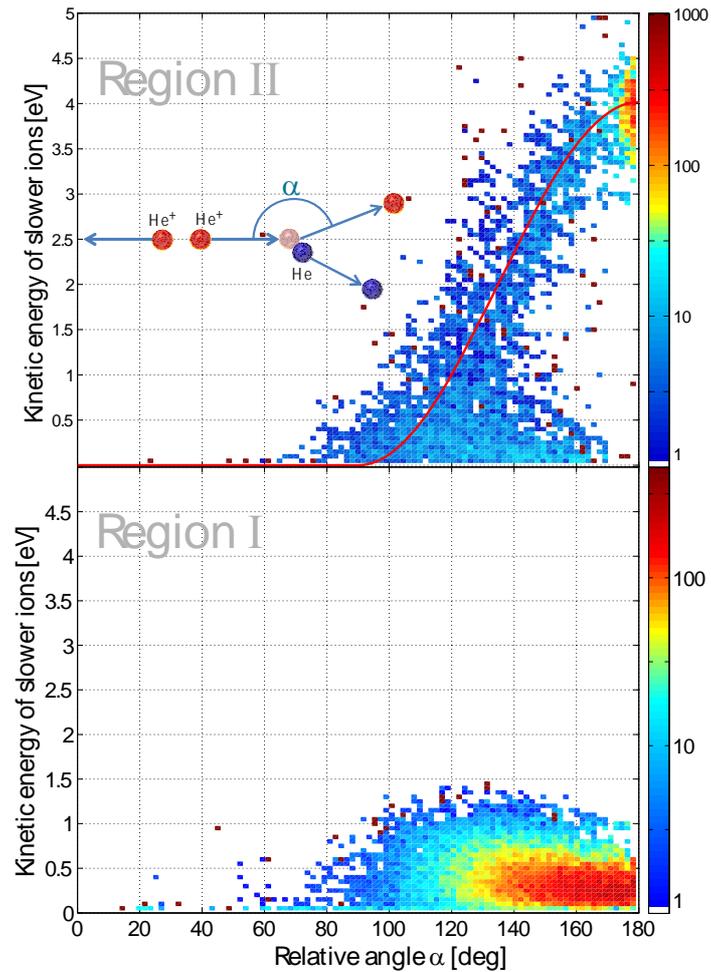

**Figure 3** Top: Kinetic energy of the slower ionic fragments from region II versus relative angle between the momentum vectors of both fragments. The simulated, red curve indicates the energy relation for an ideal elastic scattering of a helium ion from a helium atom. Bottom: Kinetic energy of the slower ionic fragments from region I relative to the angle between the momentum vectors of both fragments.

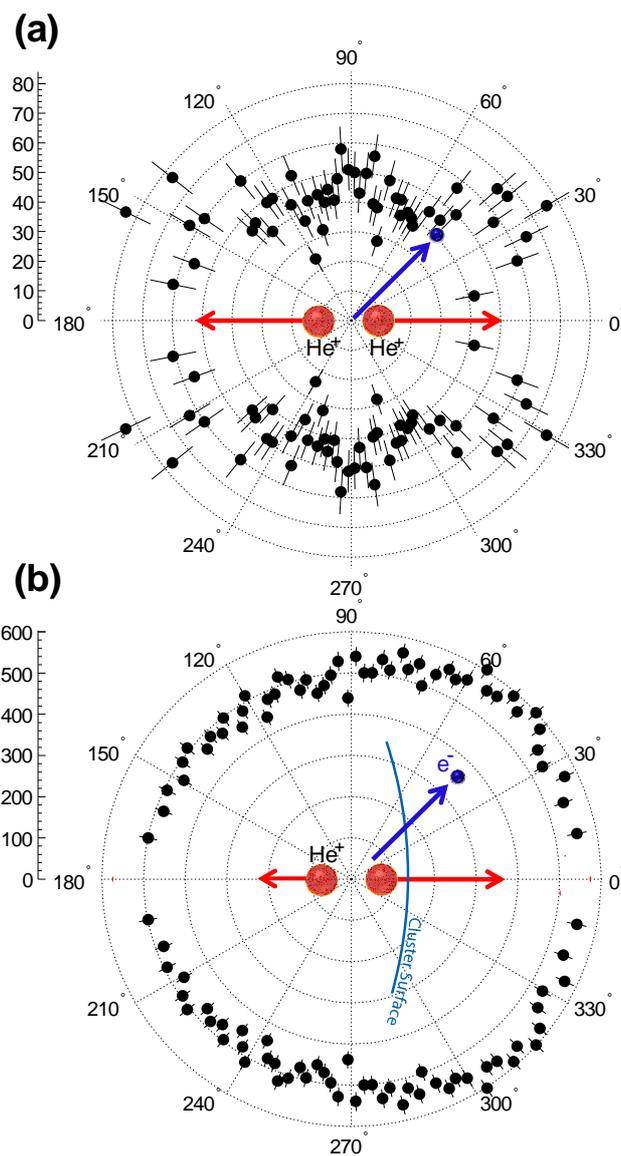

**Figure 4** Angular distribution of the ICD electron in the molecular frame of dissociating helium dimers (a) and in helium clusters in region I (b). The faster ion is emitted towards 0°.